\documentclass[12pt,preprint]{aastex}
\usepackage{emulateapj5}

\shorttitle{Radio and $\gamma$-ray emissions from pulsars: possible
observational tests} \shortauthors{Qiao et al.}

\input psfig.sty
\begin{document}

\title{Radio and $\gamma$-ray emissions from pulsars: possible observational tests}
\author{G.J. Qiao$^{1}$, K.J.Lee$^{1}$, H.G.Wang$^{2}$, R.X.Xu$^{1}$\& B.Zhang$^{3}$}
\affil{1.Astronomy Dept., School of Physics, Peking University,
           Beijing 100871, China. gjn@pku.edu.cn
2.Center for astrophysics, Guangzhou University, Guangzhou 51006,
China, cosmic008@263.net. 3.Department of Physics, University of
Nevada, Las Vegas, Nevada 89154, bzhang@physics.unlv.edu}

\begin{abstract}
Many models for the pulsar radio and $\gamma$-ray emissions have
been developed. The tests for these models using observational data
are very important. Tests for the pulsar radio emission models using
frequency-altitude relation are presented in this paper. In the
radio band, the mean pulse profiles evolve with observing
frequencies. There are various styles of pulsar profile - frequency
evolutions (which we call as "beam evolution" figure), e.g. some
pulsars  show that mean pulse profiles are wider and core emission
is higher at higher frequencies than that at lower frequencies, but
some other pulsars show completely the contrary results. We show
that all these "beam evolution" figures can be understood by the
Inverse Compton Scattering(ICS) model (see Qiao at al.2001 also). An
important observing test is that, for a certain observing frequency
different emission components are radiated from the different
heights. For the $\gamma$-ray pulsars, the geometrical method (Wang
et al. 2006) can be used to diagnose the radiation location for the
$\gamma$-ray radiation. As an example, Wang et al. (2006) constrain
the $\gamma$-ray radiation location of PSR B1055-52 to be the place
near the null charge surface. Here we show that Wang's result
matches the proposed radiation locations by the annular gap model as
well as the outer gap models.
\end{abstract}


\section{INTRODUCTION}
Rotation-powered pulsars are excellent laboratories for studying
particle acceleration as well as fundamental physics (Harding,2007).
There are several kinds of pulsar radio and $\gamma$-ray radiation
models have been published. For radio emission of pulsars, most of
them are based on the vacuum inner gap assumptions suggested by
Rudermen \& Sutherland(1975). Qiao \& Lin(1998) proposed that the
inner gap sparking would produce low frequency wave and these low
energy photons are inverse Compton scattered by the secondary
particles produced in the pair cascades, and the up-scattered radio
photons provide the observed radio emission from the pulsar. Gil \&
Sendyk(2000) emphasized that spark discharges do not occur at random
positions, instead, sparks should tend to operate in well determined
preferred regions. For the $\gamma$-ray emission form pulsars, both
polar cap and outer gap models have been suggested (e.g. Harding
1981; Harding 2007; Zhao et al. 1989; Lu \& Shi 1990; Cheng et al.
1986, 2000; Romani 1996, 2002; Hirotani 2000).

Observations are the bases for any theory. How can we take a check
for so many theories? Here we present two tests for the radiation
models basing on the observations. The first test use the well known
mean pulsar profile - frequency relation, while the second test use
the time delay between radio and high energy radiation. For the
$\gamma$-ray pulsars, the geometrical method (Wang et al. 2006) can
be used to diagnose the radiation location for the $\gamma$-ray
radiation.

\section{Check the radio emission models using "beam evolution" effects}

The typical pulsar classification are established by Rankin (1983,
1993), Manchester \& Johnston (1995), Kramer et al. (1998) and
Xilouris et al. (1998). Gil \& Sendyk (2000) interpret different
types of pulsars in the $P$-$\dot P$ diagram as having different
numbers of sparks. Qiao et al.(2001) suggest a different pulsar
classification scheme using the pulse profile evolution versus
observed frequencies relation, where the following three points are
considered. First, how the number of components in the integrated
pulsar profile changes with different observed frequencies. Second,
how the phase (in fact the radiation altitude) of there components
change with different observed frequencies. Third, how the intensity
of the components change with observing frequencies.

There are very different evolution types. For example, the PSR
B1933+16 shows the ``single''-to-triple profile evolution with
increasing frequency (Figure 1); furthermore, the separation between
the two "shoulders" of the triple profile gets wider at higher
frequencies (Qiao et al. 2001). But pulse profiles of PPSR B1237+25
are much wider than that in high frequencies. In other words, the
pulse profiles become wider at high frequencies for PSR B1933+16;
but the story for PSR B1237+25 are completely different. These
various evolution styles need to be understand theoretically.

It should be noted that these various pulsar profile - observing
frequency evolution relation ("Beam evolution") can be well
reproduced by the inverse Compton scattering(ICS) model(Qiao et al.
2001). Here we take PSR B2111+46 as an example (Figure 2), of which
the "Beam evolution"  are very complex. For the two inner conal
components, the phase separation increase, when observing frequency
gets higher. For the two outer conal components, the phase
separation decrease, when observing frequency gets higher. This is
an obvious challenge for any pulsar radio radiation models.

Recently Zhang et al. (2007) analyzed the pulse profile evolution
for PSR B2111+46 using data at seven frequency band, the radiation
heights for emission components are also calculated. It shows that
different emission components, even if at the same observing
frequency, come from different heights. This is probably a common
phenomena and challenge for any emission models. Zhang et al. (2007)
also shows that the ICS model for pulsar radio emission can explain
these results well (See Figure 2 for details).

In the next section we will review the recent result of constraining
(Wang et al. 2006 ) the $\gamma$-ray radiation models pulsars. It is
found that only the inner annular gap model and outer gap model
match the observation.

\section{To check radio \& $\gamma$-ray emission models: the
multi-band determined locations}

Observational constraints on the radio and $\gamma$ -ray emission
regions of PSR B1055-52 using both radio and $\gamma$-ray data are
given by Wang et al.(2006), we call it ``the multi- beam
constrains'', which gives allowable radio and $\gamma$-ray radiation
locations from observational point of view. Using these results one
can check related pulsar radiation theories.

In this method, the geometrical parameters, such as view angle
$\zeta$ as well as inclination angle $\alpha$ are obtained from high
quality radio linear polarization data. For PSR B1055-52 the
best-fitting value are $\alpha=74.7^{\circ}$ and the view angle
$\zeta=113.4^{\circ}$ (Lyne \& Manchester 1988 and van Ommen et al.
1997). Given these parameters, the radio and $\gamma$-ray emission
radiation locations can be constrained by fitting the observed pulse
widths of radio and $\gamma$ -ray pulses profile and the phase delay
between radio and $\gamma$-ray pulse. It is clearly that this method
determines radiation location in pure geometrical ways and do not
involve any assumption about radiation models. The detailed
calculations can be found in the paper of Wang et al.(2006).

The phase-aligned radio profile and $\gamma$ -ray light curve of
B1055-52 (Thompson et al. 1999) are shown in Figure 3. The result of
Wang et al. are present graphically in Figure 4 and 5. Radio data
suggests that the view angle of PSR B1055-52 are greater than
$90^{\circ}$. Compared with this, the polar cap model, which
suggests that the $\gamma$ -ray emission altitudes are about several
times of the stellar radius, can not explain such view angle greater
than $90^{\circ}$. So the polar cap model is not a favorable model
for PSR B1055-52. The only radiation region that allow a view angle
greater than $90^{\circ}$ are at places beyond the null charge
surface (NCS). In this way only the outer gap model(Zhang \& Cheng,
1997) or the annular model (Qiao et al.2004a,b; 2007) can be the
candidate model of explaining the $\gamma$-radiation form PSR
B1055-52.

\section{Discussions and conclusions}
How can we test pulsar radiation theories? Available data on the
mean pulse profiles of pulsars and their polarization had been
analyzed to determine the two-dimension morphology for pulsar radio
beams (Rankin,1983,1993; Lyne \& Manchester,1998). The morphology of
radiation beams presents very important observational test to any
pulsar radiation theories. Some important results are as follows:

1. Although whether the pulsar radiation beam is structured or not
is still under debate (Ranking 1983, Lyne \& Manchester 1988), the
existence of core beam was confirmed. Thus the existence of core
radiation beam is an important test to radiation model. (e.g. RS
model can not have a core beam, while ICS model have a core beam, if
the impact angle is not large.)

Rankin suggest that beside core emission beam, there are two cones
(inner cone and outer cone) for some pulsars. But Lyne \& Manchester
emphasized that the observations are best described by gradual
change in emission characteristics from the core to the outer edge
of the emission beam. Using all available multi-component radio
pulse profiles for pulsars with medium to long periods and good
polarization data, Han \& Manchester(2001) have constructed a
two-dimensional image of the mean radio beam shape. They suggest
that there is core in the beam center, but in the conical region
there are some patchy beams. Qiao et al.(2001) suggest that for some
short period pulsar, there are core and one conal beam; but for
longer period pulsars, beside the core emission beam, there are two
conal beams(inner cone and outer cone). Gil \& Sendyk(2000) put some
kinds of pulsars in the $P-{\dot P}$ diagram, it can be seen that
different kind of pulsars (Multiple, Conal Double, Conal Single Core
single,Triple) thoroughly located at different region.

2. It is well known that the pulse profiles change with observing
frequencies. Thus the relation of pulse profile-observing frequency
evolution (``Beam evolution''  figure is important test to any
emission theory on pulsar radio emission (See Qiao et al.2000 and
Zhang at al. 2007).

3. The ``multi-beam constraining'' method (Wang et al. 2006) is a
useful approach to constrain the high-energy emission regions. High
quality polarization observation are needed to derive the two key
input parameters of this method, i.e. the inclination angle $\alpha$
and the view angle $\zeta$. Since more $\gamma$-ray pulsars are
expected to be discovered after the launch of the Gamma-Ray Large
Area Space Telescope (GLAST), it will offer a large sample of
pulsars for which ``multi-beam constraining'' method can be used to
test the pulsar high-energy emission theories.

{\bf Acknowledgments}

We are very grateful to Professors. R. N. Manchester, D. Lai and J.
L. Han for their valuable discussions. This work is supported by NSF
of China (10373002, 10403001, 10273001)

\clearpage

\begin{figure}

\centerline{\psfig{figure=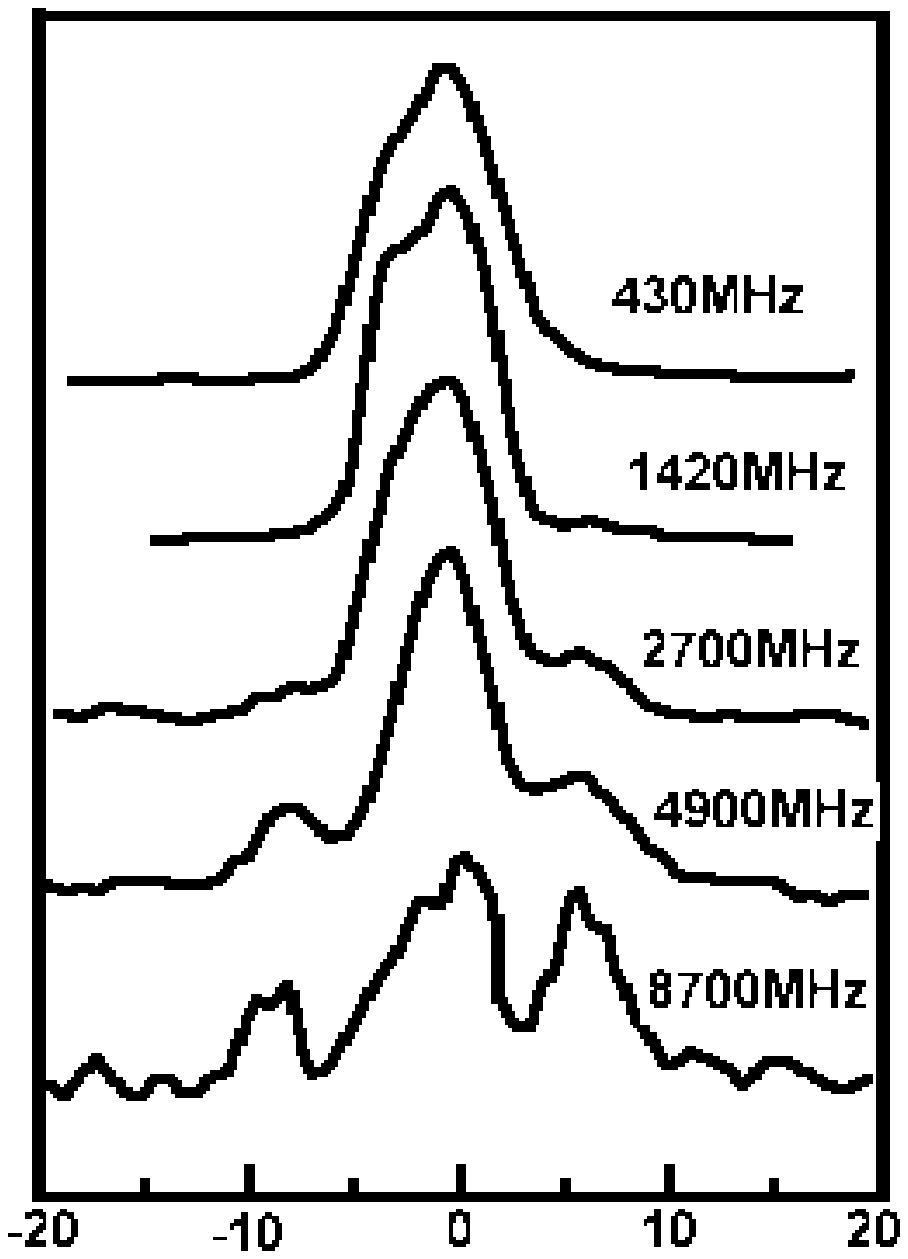,height=80mm,width=100mm,angle=0}}
\caption{The observational data of pulsar PSR B1933+16 from Lyne \&
Manchester (1988) for 1420 MHz, 2700 MHz and 8700 MHz, and from
Sieber et al. (1975) for 430 MHz and 4900 MHz. It can be seen that,
for PSR B1933+16 pulse profiles at high frequencies are wider than
that at the lower frequencies; but other pulsars such as PSR
B1237+25 and PSR B0521+21 the evolutionary behavior is opposite.}
\end{figure}




\begin{figure}

\centerline{\psfig{figure=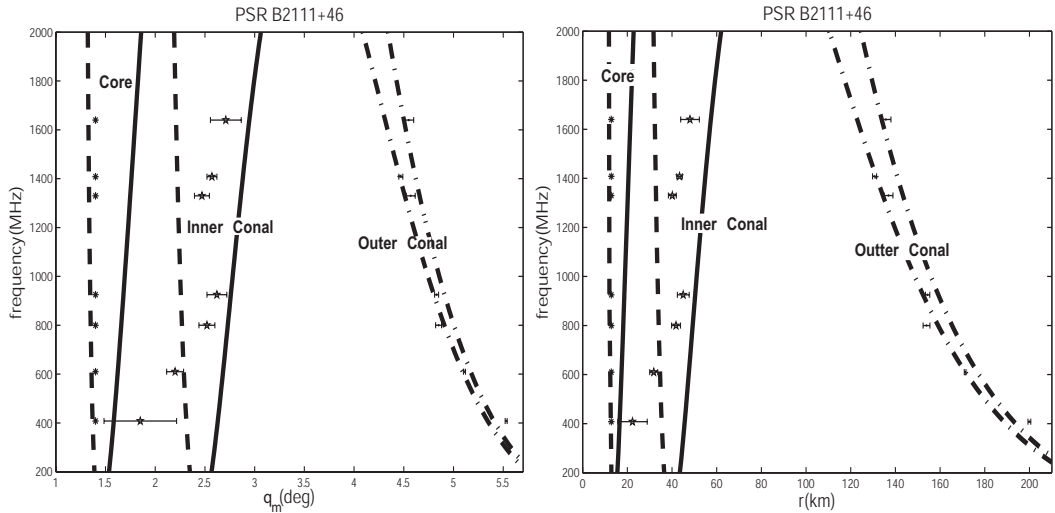,height=70mm,width=140mm,angle=0}}

\caption{A beam frequency figure of PSR B2111+46. The data points
were calculated at the field lines footed at 75 percent of the polar
cap radius by taking inclination angle $\alpha$= $9^{0}$, impact
angle $\beta$=$ 1.4^{0}$ (Rankin1993). $\Theta_{\mu}$ is the angle
between the radiation direction (in the direction of the magnetic
field) and the magnetic axis. The curves are the theoretical lines
in the ICS module calculated with the basic equations of the ICS
model. $r$ is the distance of the radiation location from the
neutron star center, which is calculated though a pure geometrical
method (Zhang et al.2007).). }
\end{figure}


\begin{figure}

\centerline{\psfig{figure=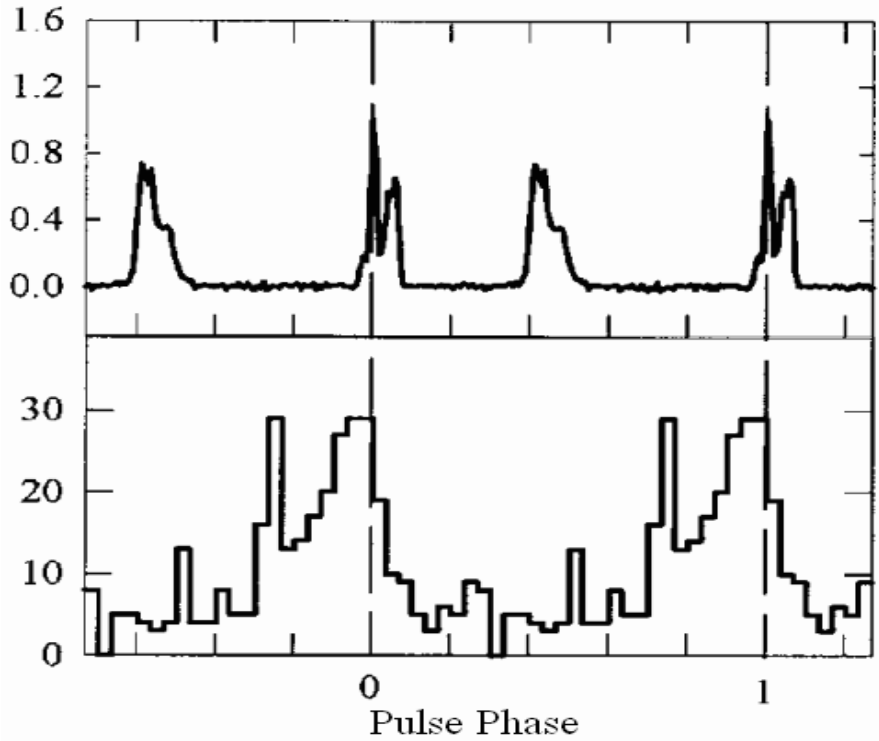,height=100mm,width=120mm,angle=0}}

\caption{The phase-aligned radio profile and $\gamma$ -ray light
curve of B1055.52 (Thompson et al. 1999). The phase of leading peak
of main pulse(MP) is chosen as a reference and is shown by a
vertical dashed line. \label{Figure4}}
\end{figure}

\begin{figure}

\centerline{\psfig{figure=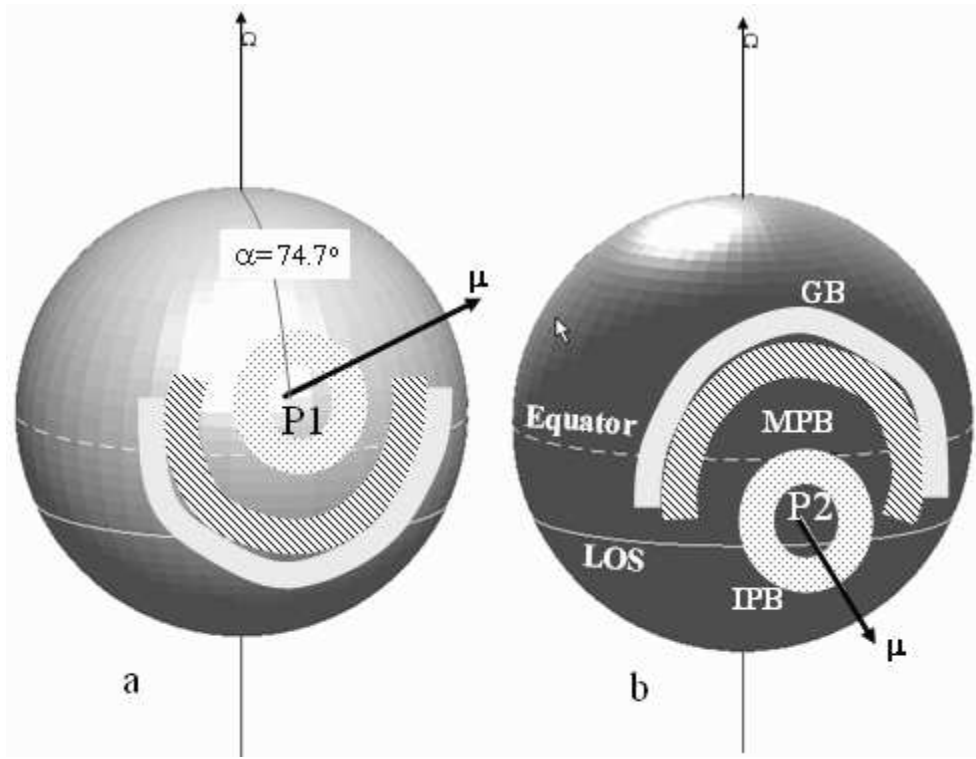,height=100mm,width=130mm,angle=0}}

\caption{A diagram for the observed central positions of the main
pulse (MP), inter pulse (IP)  and $\gamma$-ray pulse(GP).  (a) and
(b) Schematic diagrams for the geometry patterns of main pulse beam
(MPB), IP beam (IPB) and $\gamma$ -ray beam (GB) on pole 1 and pole
2, respectively. }
\end{figure}

\begin{figure}

\centerline{\psfig{figure=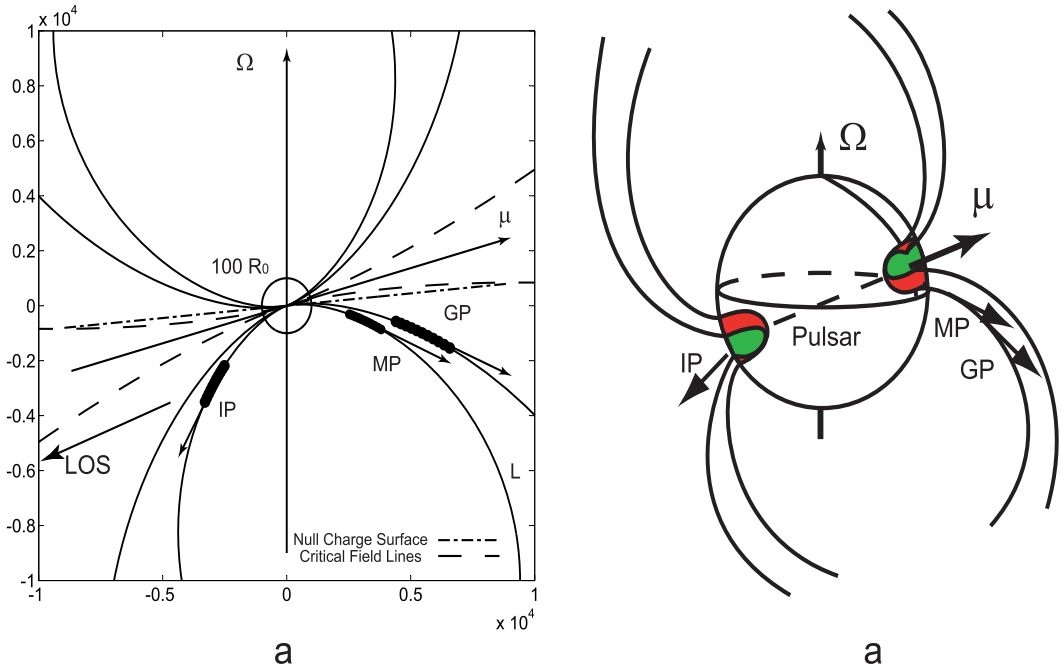,height=90mm,width=140mm,angle=0}}

\caption{(a). The 2D diagram for the observed radio main pulse beam
(MPB) on $\Omega$-$\mu$ plane. The radiation locations for the radio
inter-pulse beam (IPB) and the $\gamma$ -ray beam (GB) are also
plotted for illustrative purposes on the $\Omega$-$\mu$ plane. The
radiation direction of the IPB and GB are lower than that of the
line of the sight(LOS), so any observer can not observe it at this
plane(here it is in between of the two pulses). The null charge
surface is plotted using dot-dashed lines, while the critical
magnetic field lines are plotted using dashed lines. The direction
for the line of sight is given by the arrow labeled by `LOS'. Here
$\Omega$ is the rotational axis; $\mu$ is the magnetic axis. It is
very clear that the radiation locations stay near or beyond the null
charge surface. More important, the radiation comes from the
"unfavorable" magnetic field lines. The inclination angle
$\alpha=74.7^{\circ}$ and the view angle $\zeta=113.4^{\circ}$.
Detailed calculation for plotting this figure can be found in Wang
et al.(2006).(b). A schematic figure of the radiation location for
different emission beams. The red regions are the annular regions,
along which are the annular acceleration regions. }
\end{figure}

\end{document}